%% file: main.tex
\documentclass[lettersize,journal]{IEEEtran}

\ifCLASSOPTIONcompsoc
  \usepackage[nocompress]{cite}
\else
  \usepackage{cite}
\fi

\ifCLASSINFOpdf
\else
\fi

\usepackage[pdftex]{hyperref}

\usepackage[ruled]{algorithm2e}
\usepackage{my-tvcg}
\usepackage{epigraph}
\graphicspath{{./images/}}

\usepackage{color}
\definecolor{MainColor}{RGB}{128, 0, 128}

\hypersetup{
    unicode=true,          
    pdftoolbar=true,        
    pdfmenubar=true,        
    pdffitwindow=false,     
    pdfstartview={FitH},    
    pdftitle={Polynomial 2D Green Coordinates for High-order Cages},    
    pdfnewwindow=true,      
    colorlinks=true,       
    linkcolor=MainColor,          
    citecolor=MainColor,        
    filecolor=magenta,      
    urlcolor=MainColor,           
    breaklinks=MainColor
}

\begin{document}

\title{Polynomial 2D Biharmonic Coordinates for High-order Cages}

\author{Shibo~Liu,
        Ligang~Liu,
        Xiao-Ming~Fu
\IEEEcompsocitemizethanks
{
\IEEEcompsocthanksitem S. Liu, L. Liu, and X. Fu are with the School of Mathematical Sciences, University of Science and Technology of China.\protect\\
E-mail: aa1758926168@mail.ustc.edu.cn, lgliu@ustc.edu.cn, fuxm@ustc.edu.cn.
\IEEEcompsocthanksitem Corresponding author: Xiao-Ming~Fu.
}
}

\markboth{Journal of \LaTeX\ Class Files,~Vol.~14, No.~8, December~2019}%
{Shell \MakeLowercase{\textit{et al.}}: Bare Demo of IEEEtran.cls for Computer Society Journals}

\IEEEtitleabstractindextext{%
\begin{abstract}
We derive closed-form expressions of biharmonic coordinates for 2D high-order cages, enabling the transformation of the input polynomial curves into polynomial curves of any order.
Central to our derivation is the use of the high-order boundary element method.
We demonstrate the practicality and effectiveness of our method on various 2D deformations. In practice, users can easily manipulate the \Bezier control points to perform the desired intuitive deformation, as the biharmonic coordinates provide an enriched deformation space and encourage the alignment between the boundary cage and its interior geometry.
\end{abstract}

}

\maketitle
\IEEEdisplaynontitleabstractindextext
\IEEEpeerreviewmaketitle

\input{src/introduction}

\input{src/relatedwork}

\input{src/method}
\input{src/results}

\input{src/conclusion}

\input{src/acknowledgments}

\ifCLASSOPTIONcaptionsoff
  \newpage
\fi

\bibliographystyle{IEEEtran}
\bibliography{src/reference}
%

\begin{IEEEbiography}[{\includegraphics[width=1in,height=1.25in,clip,keepaspectratio]{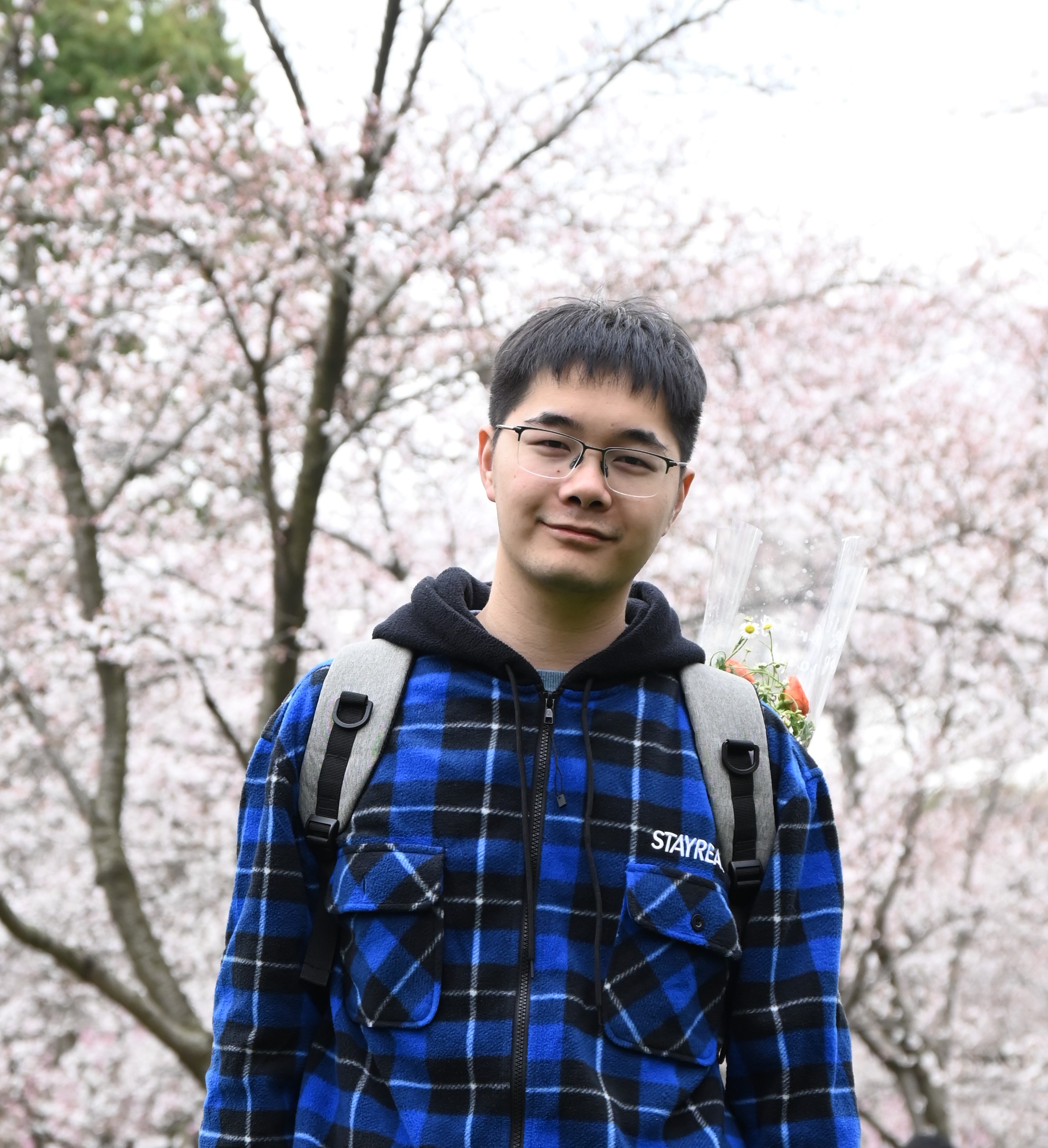}}]{Shibo Liu}
received a BSc degree in 2021 from the University of Science and Technology of China. He is currently a PhD candidate at the School of Mathematical Sciences, University of Science and Technology of China. His research interests include high-order geometric processing and 3D modeling. His research work can be found at his research website: \url{https://liu43.github.io/}.
\end{IEEEbiography}

\begin{IEEEbiography}[{\includegraphics[width=1in,height=1.25in,clip,keepaspectratio]{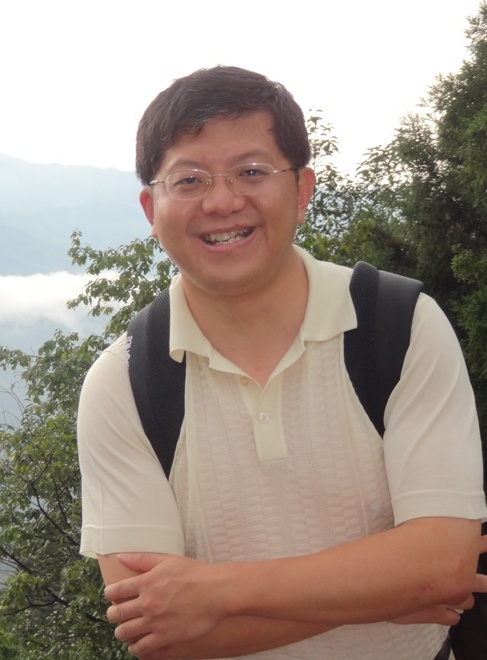}}]{Ligang Liu}
is a Professor at the School of Mathematical Sciences, University of Science and Technology of China. His research interests include computer graphics and CAD/CAE. His work on light-weight designing for fabrication at Siggraph 2013 was awarded as the first Test-of-Time Award at Siggraph 2023. \url{http://staff.ustc.edu.cn/~lgliu}.
\end{IEEEbiography}

\begin{IEEEbiography}[{\includegraphics[width=1in,height=1.25in,clip,keepaspectratio]{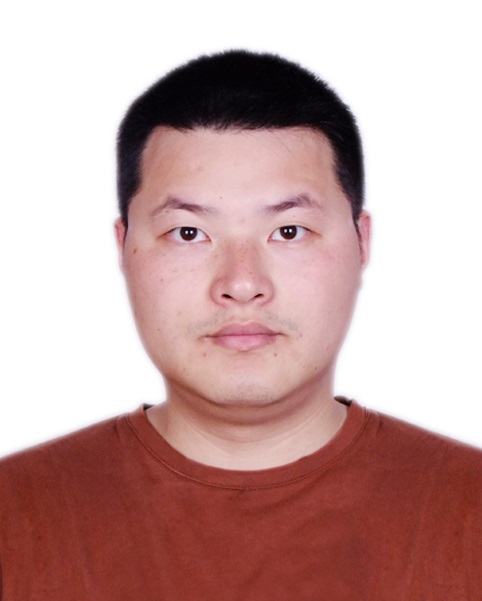}}]{Xiao-Ming Fu}
received a BSc degree in 2011 and a PhD degree in 2016 from the University of Science and Technology of China. He is an associate professor at the School of Mathematical Sciences, University of Science and Technology of China. His research interests include geometric processing and computer-aided geometric design. His research work can be found at his research website: \url{https://ustc-gcl-f.github.io/}.
\end{IEEEbiography}

\end{document}

%% file: src/introduction.tex
\section{Introduction} \label{sec:intro}

A 2D cage contains a set of curves to enclose a domain.
After specifying sparse scalar or vector values on the cage, the cage coordinate is powerful enough to determine the value everywhere in the domain.
For example, the coordinate drives the shape deformation when specifying sparse displacements for the cage or edits the color for the domain when defining colors on the cage. 

\begin{figure}[t]
  \centering
  \begin{overpic}[width=1.0\linewidth]{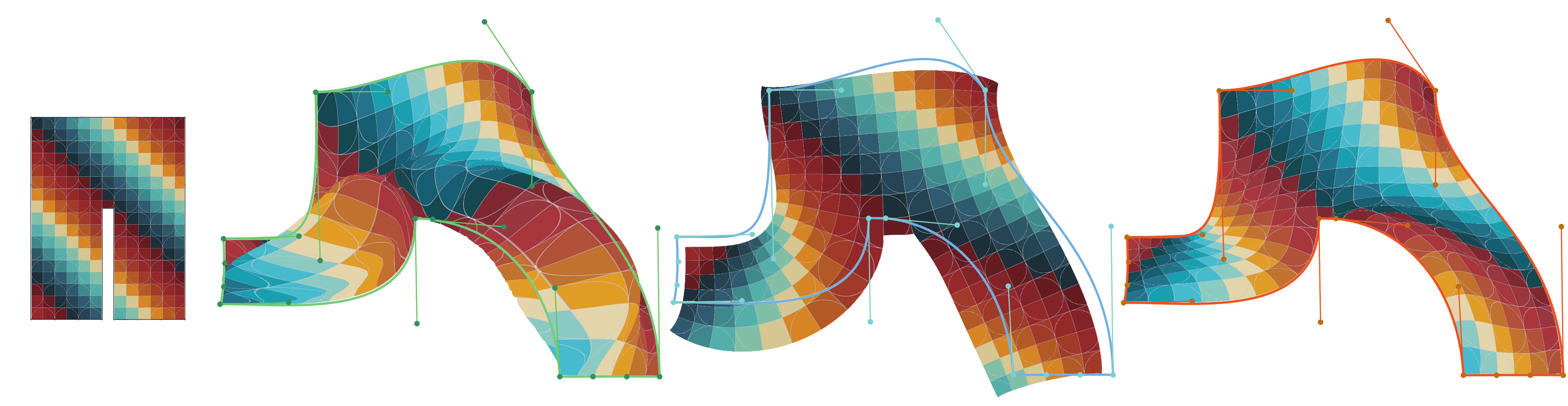}
    {
     \put(5,-1.5){\small \textbf{(a)}}
      \put(17,-1.5){\small \textbf{(b)} Cubic MVC}
       \put(49,-1.5){\small \textbf{(c)} PolyGC}
 \put(79,-1.5){\small \textbf{(d)} Ours}
    }
  \end{overpic}
  \caption{
  Comparison with Cubic MVC and PolyGC on the Pants shape with a polygonal cage (a). 
  %
(b) Cubic MVC~\cite{Li2013} interpolates the boundary but causes severe visual artifacts.
(c) PolyGC~\cite{MichelThiery2023} produces conformal deformation but misses the boundary alignment.
(d) Our coordinates provide a favorable trade-off between boundary alignment and deformation distortion control. 
  }
  \label{fig:cmp-cmvc}
\end{figure}

Many cage coordinates have been proposed and can be classified into two categories.
First, after specifying a value for each polygonal cage vertex, the first type of coordinates provides ways to interpolate an arbitrary point through a weighted combination of these values, such as mean value coordinates and their variants (MVC)~\cite{Floater2003, Hormann2006, Lipman2007, Li2013}, and harmonic coordinates~\cite{Joshi2007}. These coordinates have the property of interpolating the boundary; however, visible deformation artifacts often arise  (Fig.~\ref{fig:cmp-cmvc}~(b)).
%
Second, another type of coordinates offers control over boundary values and boundary derivatives (i.e., normals)~\cite{Lipman2008,  Hou2018, Ilbery2013, MichelThiery2023,liu2024polygc}. Most of them can control the deformation distortion to a low level. Still, the alignment between the cage and the shape boundary is poor, increasing the difficulty in editing deformations (Fig.~\ref{fig:cmp-cmvc}~(c)).

As a natural extension to the harmonic coordinates, biharmonic coordinates~\cite{Weber2012,thiery2024biharmonic} with an enriched deformation space are superior for 2D shape deformation while having comparable alignment between the cage and the shape boundary (Fig.~\ref{fig:cmp-cmvc}~(d)).
As pioneers of biharmonic coordinates, \cite{Weber2012} study closed-form coordinates for 2D polygonal cages and ~\cite{thiery2024biharmonic} derive closed-form coordinates and their derivatives for 3D triangular cages.
%

\begin{figure}[t]
		\centering
		\begin{overpic}[width=0.99\linewidth]{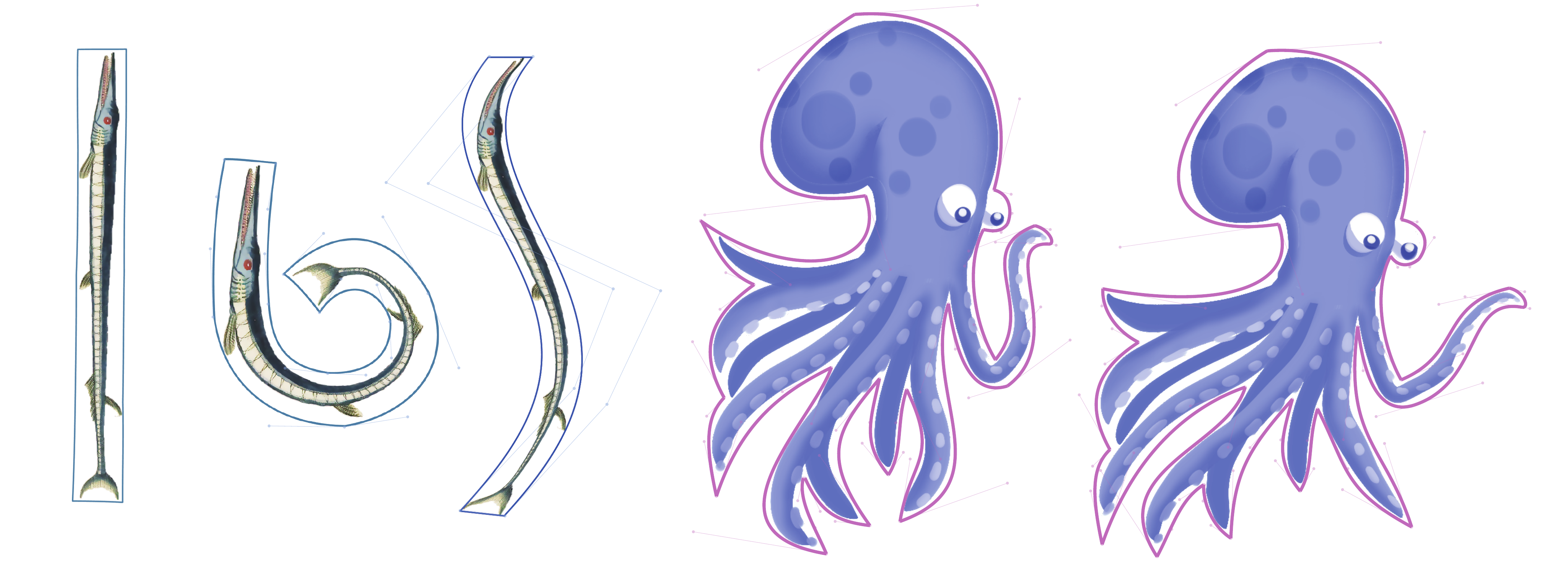}
			{
				   \put(5,-1.5){\small \textbf{(a1)}}
                   \put(18,-1.5){\small \textbf{(a2)}}
                   \put(30,-1.5){\small \textbf{(a3)}}
                      \put(56,-1.5){\small \textbf{(b1)}}
                       \put(83,-1.5){\small \textbf{(b2)}}
		    }
		\end{overpic}
		 \vspace{-2mm}
		 \caption{
         Deformation using our coordinates.
		Given an input image and a polygonal cage (a1), our coordinates enable deformations to be high-order cages with polynomial curves of different orders (3 in (a2) and 4 in (a3)).
        Moreover, we can transform a cubic input high-order cage (b1) into cubic (b2).
			 }
		 \label{fig:teaser}
		 \end{figure}
         
\begin{figure}[t]
  \centering
  \begin{overpic}[width=1.0\linewidth]{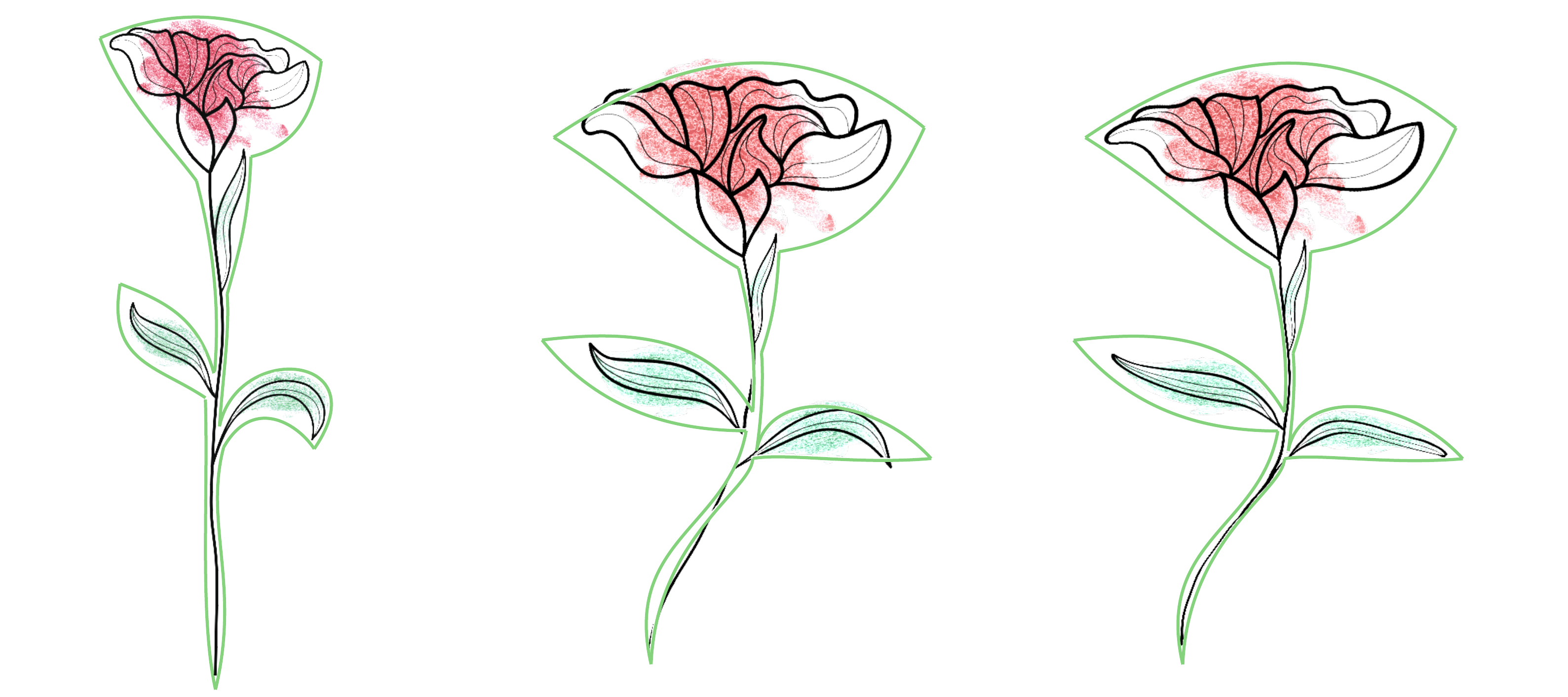}
    {
    \put(7,-1.5){\small \textbf{(a)} Rest pose}
    \put(38,-1.5){\small \textbf{(b)} CurvedGC}
 \put(75,-1.5){\small \textbf{(c)} Ours}
    }
  \end{overpic}
  \caption{
  Given an input shape enclosed by a cubic high-order cage (a), our coordinates (c) lead to better alignment between the cage and the shape boundary than~\cite{liu2024polygc} (b).
  }
  \label{fig:cmp-CurvedGC}
\end{figure}

\begin{figure}[t]
  \centering
  \begin{overpic}[width=1.0\linewidth]{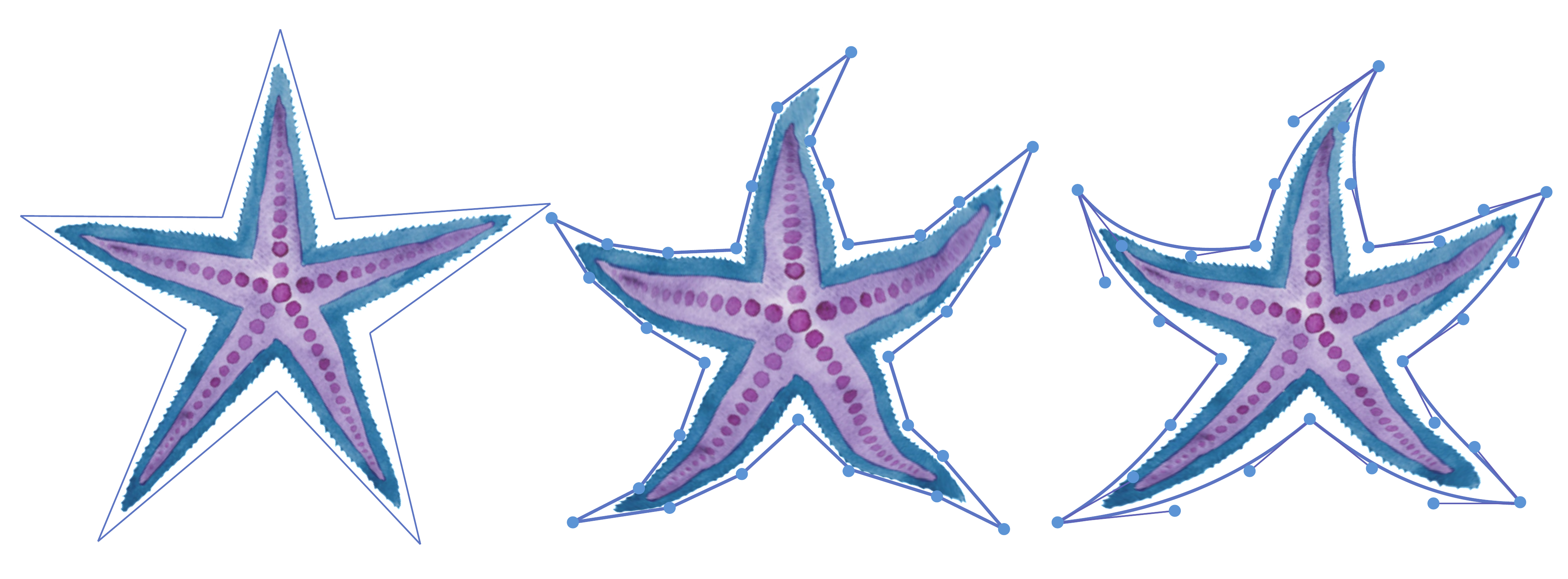}
    {
     \put(7,-1.5){\small \textbf{(a)} Rest pose}
    \put(39,-1.5){\small \textbf{(b)} Biharmonic}
 \put(68,-1.5){\small \textbf{(c)} Cubic biharmonic}
    }
  \end{overpic}
  \caption{
  Using the same number of control points, deforming the curved cage with 10 cubic curves (c) achieves smoother and more intuitive editing than deforming the polygonal cage having 40 straight segments (b).
  }
  \label{fig:linear-vs-curved}
\end{figure}

Unlike polygonal cages, high-order cages, consisting of polynomial segments, provide a natural way to control the tangential stretch and curvature along the deformation cages (Fig.~\ref{fig:linear-vs-curved}).
Moreover, the high-order cage can approximate the input shape better than the linear cage.
Due to these advantages, deriving closed-form coordinates for curved cages has attracted great attention.
Several coordinates, such as Cubic MVC~\cite{Li2013} and polynomial Green coordinates~\cite{MichelThiery2023} (PolyGC), and polynomial cauchy coordinates~\cite{Lin2024PolynomialCC}, transform linear cages to curved cages.
They need intermediate straight cages to realize the deformation from high-order cages to high-order cages.
\cite{liu2024polygc} extend GC to be suitable for high-order input cages, called \emph{CurvedGC}.
%
However, 2D biharmonic coordinates for high-order cages have remained unexplored. 

In this paper, we provide the missing coordinates, i.e., derive closed-form expressions of biharmonic coordinates for 2D high-order cages.
The core of our derivation is applying the high-order boundary element method and computing the integral analytically.
Obviously, our coordinates inherit the strengths of biharmonic coordinates and high-order cages as follows (Fig.~\ref{fig:cmp-CurvedGC}).
First, allowing control over boundary values and derivatives to mitigate deformation artifacts.
Second, the deformed shape boundary respects the boundary cage, making the deformation conform to the user's intention and lowering the user's manipulation difficulty.
Third, the user has more degrees of freedom to manage deformation shapes due to the high-order cages.
Our cage-based deformation experiments demonstrate these benefits.

%% file: src/relatedwork.tex
\section{Related work} \label{sec:related}

\paragraph{Interpolating coordinates}
Many interpolating barycentric coordinates are proposed to give users precise and intuitive shape control, i.e., make the deformed shape boundary follow the cage tightly.
Many coordinates begin with the properties of harmonic functions.
The mean value theorem of harmonic functions is used to derive the mean value coordinates (MVCs) with closed forms~\cite{Floater2003,Hormann2006,Ju2005}.
MVCs are suitable for multiresolution deformation frameworks due to their closed-form expressions~\cite{Huang2006,Farbman2009}. 
A well-known weakness of MVC is that negative coordinates will appear, leading to unnatural deformation behaviors.
To solve issues, positive mean-value coordinates~\cite{Lipman2007} and harmonic coordinates~\cite{Joshi2007} are introduced, but they lack closed-form solutions.
%
Poisson Coordinates~\cite{li2012poisson} offers better accuracy in representing harmonic functions.
There are many other coordinates designed from other principles. 
Maximum entropy coordinates~\cite{Hormann2008Maximum} and maximum likelihood coordinates~\cite{Chang2023Maximum} are linked to information theory. 
Local barycentric coordinates~\cite{Zhang2014Local} focus on locality.
Variational barycentric coordinates~\cite{Dodik2023Variational} are computed as the minimizer of the total variation. 
These coordinates require an optimization process and do not possess closed-form expressions.
Although accurate shape control near the cage can be provided by performing interpolation on the cage via any of these coordinates, unfortunately, great deformation distortion often arises.

%

\paragraph{Non-interpolating coordinates}
Cage coordinates, which provide control over boundary values and boundary derivatives (i.e., normals), are developed to mitigate the deformation artifacts.
Based on Green's third identity of harmonic functions, \cite{Lipman2008} propose Green coordinates to achieve conformal deformations in 2D and quasi-conformal mappings in 3D.
\cite{Weber2009} use the Cauchy integral formula for holomorphic functions to derive Cauchy coordinates, which are shown to be equivalent to 2D Green coordinates. 
These two coordinates sacrifice the precise shape control.
The boundary integral formulation of linear elasticity is used to introduce Somigliana coordinates~\cite{chen2023somigliana}, which enable volume control of cage deformation.
The closed-form biharmonic coordinates in 2D~\cite{Weber2012} and in 3D~\cite{thiery2024biharmonic} have an enriched deformation space to achieve intuitive smooth deformations while possessing a small misalignment between the cage and the shape boundary.

\paragraph{Extension to high-order cages}
Extending existing coordinates to high-order cages receives much attention as such cages offer a powerful way to control the tangential stretch and curvature along the deformation cages.
Cubic mean-value coordinates~\cite{Li2013} (Cubic MVC) are extensions of MVCs to allow the deformation of polygonal cage edges into cubic curves. 
As an extension of Green coordinates, polynomial 2D Green coordinates~\cite{MichelThiery2023} have closed-form solutions to allow transforming linear cage segments into polynomial curves of any degree. 
\cite{Lin2024PolynomialCC} extend Cauchy coordinates for the same purpose and further generalize them for application to high-order cages.
To realize the deformations from high-order cages to high-order cages, these coordinates need an intermediate straight cage, which serves as the actual input configuration. 
\cite{liu2024polygc} further derive closed-form Green coordinates for 2D high-order cages so that the input polynomial cage segments can be deformed to polynomial curves of any order.
Our goal is to drive closed-form biharmonic coordinates for 2D high-order cages, similar to~\cite{liu2024polygc}.

\paragraph{Boundary element method in graphics}
When solving a linear partial differential equation (PDE) with boundary conditions defined on the domain $\Omega $, the use of the Method of the Boundary Element Method (BEM) provides a way to solve the PDE with only a boundary discretization. BEM was first introduced in the
graphics community for real time deformable objects simulations~\cite{James1999ArtDefoAR}, followed by ocean wave animation ~\cite{Schreck2019}, and surface only liquids simulation \cite{Da2016SurfaceonlyL}. Beyond simulations, BEM has proven valuable in geometry processing tasks such as computing fields of orthogonal directions within a volume~\cite{Solomon2017BoundaryEO}, constructing planar harmonic mappings \cite{Levi2016OnTC} and diffusion curves~\cite{Bang2023}. Recently, fast BEM solver \cite{Chen2024LightningfastMO} greatly improves the efficiency of BEM solves for large-scale problems.


%

%

%% file: src/method.tex
\section{method}\label{sec:method}
We review BEM and its application to derive biharmonic coordinates in Section~\ref{sec:revisit}, then introduce our closed-form coordinates for high-order cages based on high-order BEM in Section~\ref{sec:HighOrderBEM}, and finally discuss deformation controls in Section~\ref{sec:deformation}.

\subsection{Boundary element method}\label{sec:revisit}
\paragraph{BEM overview}
To solve a partial differential equation with BEM, it generally has the following steps:
\begin{enumerate}
    \item The differential equation within the domain is converted into an integral equation on the boundary. 
    \item The boundary is then discretized into finite-sized boundary elements.
    \item The boundary integral equation is transformed into an algebraic equation.
    \item Solving the algebraic equation to obtain the solution of the original partial differential equation.
\end{enumerate}
Next, we show that applying BEM to obtain closed-form biharmonic coordinates for linear cages in 2D~\cite{Weber2012} and 3D~\cite{thiery2024biharmonic}.

\paragraph{Biharmonic Dirichlet problem}
Biharmonic coordinates are derived as the solution to the biharmonic Dirichlet problem in a bounded domain $\Omega$ (refer to Equation (4) in \cite{Weber2012}):
\begin{equation}\label{equ:bihPDE}
\begin{aligned}
    &\Delta^2 f(\eta) = 0, \quad \eta \in \Omega, \\
    &f(\xi) = g_1(\xi), \quad \xi \in \partial \Omega, \\
    &\frac{\partial f}{\partial \mn_{\xi}}(\xi) = g_2(\xi), \quad \xi \in \partial \Omega,
\end{aligned}
\end{equation}
where $f$ is a biharmonic function, \( g_1 \) and \( g_2 \) are the prescribed Dirichlet and Neumann boundary conditions, respectively, and \( \mn_{\xi} \) is the normal of the cage \( \partial \Omega \) at \( \xi \). 

\paragraph{Boundary integral equations}
Using Green's theorem and the fundamental solution, the partial differential equation~\eqref{equ:bihPDE} is transformed into the following boundary integral equations:
\begin{equation}\label{equ:BIE1}
\begin{aligned}
f(\eta) &=
\int\limits_{\xi \in \partial \Omega}
f(\xi) \frac{\partial G_1}{\partial \mn_{\xi}}(\xi, \eta)d\xi -
\int\limits_{\xi \in \partial \Omega}
G_1(\xi, \eta) \frac{\partial f(\xi)}{\partial \mn_{\xi}}d\xi \\
+ &\int\limits_{\xi \in \partial \Omega}
\Delta f(\xi) \frac{\partial G_2}{\partial \mn_{\xi}}(\xi, \eta)d\xi -
\int\limits_{\xi \in \partial \Omega}
G_2(\xi, \eta) \frac{\partial \Delta f(\xi)}{\partial \mn_{\xi}}d\xi,
\end{aligned}
\end{equation}

\begin{equation}\label{equ:BIE2}
\begin{aligned}
\Delta f(\eta) =
\int\limits_{\xi \in \partial \Omega}
\Delta f(\xi) \frac{\partial G_1}{\partial \mn_{\xi}}(\xi, \eta)d\xi -
\int\limits_{\xi \in \partial \Omega}
G_1(\xi, \eta) \frac{\partial \Delta f(\xi)}{\partial \mn_{\xi}}d\xi,
\end{aligned}
\end{equation}
where $G_1$ and $G_2$ are the fundamental solutions of the 2D harmonic and biharmonic equations, respectively. Specifically, these are given by $G_1(\xi,\eta)=-\frac{1}{2 \pi}\ln\|\xi-\eta\|$ and $G_2(\xi,\eta)=-\frac{\|\xi-\eta\|^2}{8 \pi}(\ln\|\xi-\eta\|-1).$
In the integral equations, since \( f \) and \( \partial f/\partial \mn \) are given in~\eqref{equ:bihPDE}, the unknown functions only contain \( \Delta f \) and \( \partial \Delta f/\partial \mn \). 
Once we have solved for all the unknown function values of \( \Delta f \) and \( \partial \Delta f/\partial \mn \) on the boundary, substituting them into~\eqref{equ:BIE1} yields the value of $f$ at any point within the domain $\Omega$, which is the solution to~\eqref{equ:bihPDE}.
%
To solve for \(\Delta f\) and \(\partial \Delta f/\partial \mn\) on the boundary, \(\eta\) must be moved to the boundary (denoted as \(\eta^{\text{b}}\)) in~\eqref{equ:BIE1} and~\eqref{equ:BIE2}. 

\paragraph{Biharmonic Coordinates for linear cages}
In general, we discretize the integral equations to obtain algebraic equations that are further solved to obtain the solution numerically.
The discretization process consists of two steps. 
First, the boundary is divided into finite-sized boundary elements. For 2D polygonal cages~\cite{Weber2012} and 3D triangular cages~\cite{thiery2024biharmonic}, the boundary is piecewise linear and is naturally discretized into piecewise linear elements. 
Second, the functions defined on the boundary are approximated within each boundary element using interpolation. 
Within each element, \(f\) and \( \partial f/\partial \mn\) are represented using piecewise linear and piecewise constant functions, respectively. The unknowns \(\Delta f\) and \(\partial \Delta f/\partial \mn\) are also approximated using piecewise linear and piecewise constant functions. 
By representing the functions in terms of their values at the nodes of the boundary elements, an algebraic system of equations is established based on the evaluation of \(\eta'\) at different boundary points.
To solve the system of algebraic equations, \cite{Weber2012} substitute the algebraic constraints~\eqref{equ:BIE2} into the equations~\eqref{equ:BIE1} to form a linear system that is further solved. 
In 3D, \cite{thiery2024biharmonic} simultaneously minimize the two constraints, resulting in better-behaved deformations.

\begin{figure}[t]
  \centering
  \begin{overpic}[width=0.99\linewidth]{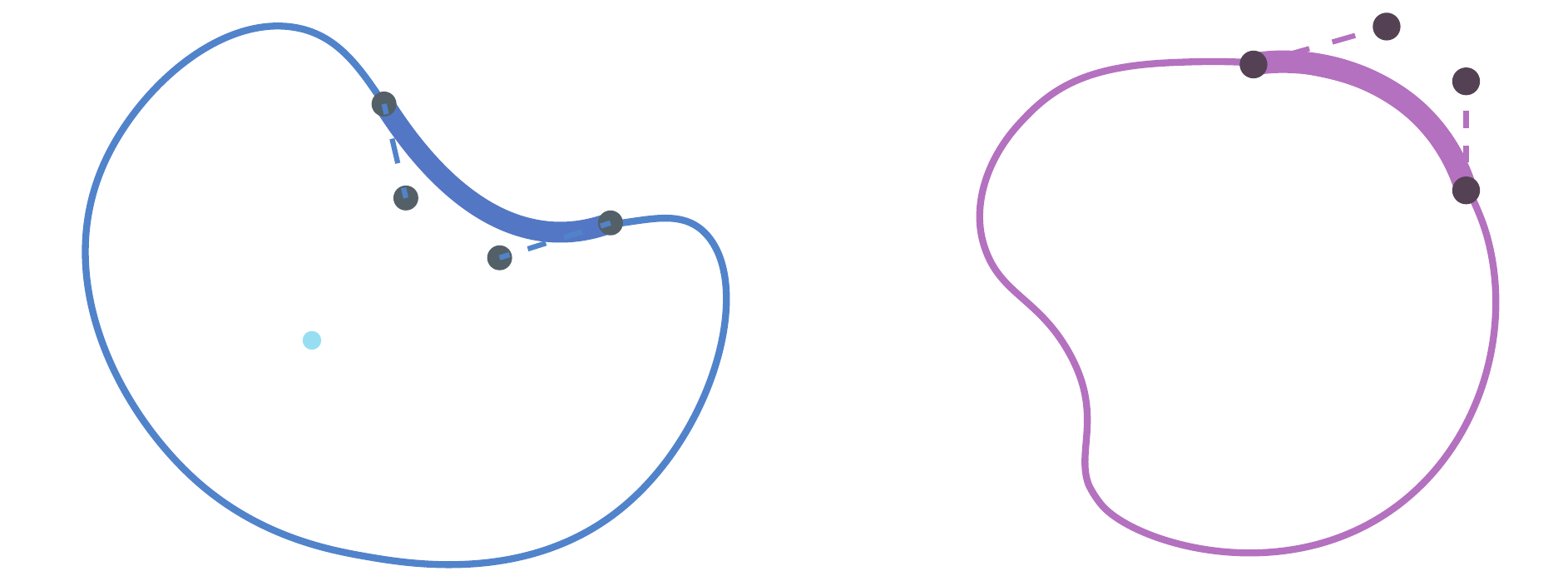}
    {
 \put(17,12){\small $\eta$}
  \put(29,29){\small $\mc_i$}
   \put(87,27){\small $\mg_1$}
    }
  \end{overpic}
  \vspace{-3mm}
  \caption{
  High-order boundary element method. The curved boundary is composed of a series of cubic elements $\mc_i$, on which $\mg_1$ is a cubic polynomial function.
  }
  \label{fig:schematic}
\end{figure}

\subsection{Polynomial Biharmonic Coordinates}\label{sec:HighOrderBEM}



To achieve deformations from curved edges to curved edges, we apply the high-order boundary element method~\cite{Fischer1999ApplicationOH,Frijns2000ImprovingTA,gravenkamp2019high}, which requires high-order elements and shape functions.

\subsubsection{Reformulating boundary integral equations}\label{sec:Reformulating boundary integral equations}
\paragraph{High-order elements}
The input 2D cage consists of $N_c$ polynomial curves. 
These curves are independent in the boundary integral equations.
The $i$-th boundary element is a polynomial curve of order $m$, parametrized as \( \mc_i(t): [0, 1] \to \mathbb{R}^2 \).
This curve $\mc_i(t)$ can be represented as an \(m\)-th order \Bezier curve:
\begin{equation}
    \mc_i(t)=\sum_{j=0}^m \mc_{j}^i B_j^m(t),
\end{equation}
where \( B_j^m \) denotes the Bernstein basis function, and \( \mc_{j}^i \) represents the corresponding control point.

\paragraph{High-order boundary conditions}
Typically, we assume that \(\mg_1\) is an \(n\)-th order polynomial function on the element $\mc_i(t)$:
\begin{equation}\label{equ:g}
\begin{aligned}
    \mg_1(t)=\sum_{j=0}^n \mg_{1,j}^i B_j^n(t), \,\, \mg_2(t)=s_i\overbrace{\frac{\|\mg_1'(t)\|}{\|\mc_i'(t)\|}}^{\sigma} \overbrace{\frac{\mg_1'(t)^{\perp}}{\|\mg_1'(t)^{\perp}\|}}^{\bar{\mn}}=s_i\frac{\mg_1'(t)^{\perp}}{\|\mc_i'(t)\|},
    \end{aligned}
\end{equation}
where \( \mg_{1,j}^i \) represents the control point and \((a, b)^\perp = (b, -a)\). Here, we impose Neumann boundary conditions using the scaling factor \(\sigma\) multiplied by the unit normal. Simply setting \(\sigma\) to 1 leads to the emergence of an undesirable term \(\|\mg_1'(t)^\perp\|\). Instead, we utilize \(\sigma\) the same as in Green coordinates. This approach ensures boundary behavior similar to~\cite{MichelThiery2023}. Additionally, we introduce a scaling factor \(s_i\), which is typically set to 1. More complex Neumann boundary condition configurations about $s_i$ are discussed in Section~\ref{sec:deformation}.
Since $\mg_1(t)$ implies that the deformed edge of the cage is an \(n\)-th degree polynomial curve, we choose \(n \geq m\) to ensure the deformation continuity.

\paragraph{High-order shape functions}
We use high-order shape functions to approximate the unknown functions \(\Delta f\) and \(\partial \Delta f/\partial \mn\) on the boundary.
Typically, we approximate \(\Delta f\) using a \(k\)-th order polynomial and \(\partial \Delta f/\partial \mn\) using a \((k-1)\)-th order polynomial on $\mc_i(t)$:
\begin{equation}\label{equ:Df}
\begin{aligned}
    \Delta f(t)=\sum_{j=0}^k \mh_{1,j}^i B_j^k(t), \frac{\partial \Delta f}{\partial \mn}(t)=\sum_{j=0}^{k-1} \mh_{2,j}^i B_j^{k-1}(t),
    \end{aligned}
\end{equation}
where \( \mh_{1,j}^i \) and \( \mh_{2,j}^i \) are the control points.
We choose \(k \geq n\) to ensure that affine transformations can be reproduced.

\paragraph{Reformulation}
Note that $\frac{\partial G}{\partial \mn}=\nabla_1 G(\mc(t),\eta)\cdot \mn_{c(t)}$ and $\mn_{\mc(t)} d\xi=\mn_{\mc(t)}\|\mc'(t)\|dt=\mc'(t)^{\perp}dt$. 
Substituting \eqref{equ:g} and \eqref{equ:Df} into \eqref{equ:BIE1} and \eqref{equ:BIE2} leads to the discretized boundary integral equations: 
\begin{equation}\label{equ:dis_Int1}
\begin{aligned}
    f(\eta) = \sum_{i=1}^{N_c} &\left(\sum_{j =0}^n \phi_{i,j}(\eta) \mg_{1,j}^i +  \sum_{j =0}^{n-1}\psi_{i,j}(\eta) s_i (\mg_{2,j}^{i})^{\perp} \right.\\
&+\left. \sum_{j =0}^k \tilde{\phi}_{i,j}(\eta) \mh_{1,j}^i +  \sum_{j =0}^{k-1}\tilde{\psi}_{i,j}(\eta) (\mh_{2,j}^{i})^{\perp}\right),
\end{aligned}
\end{equation}
\begin{equation}\label{equ:dis_Int2}
   \Delta f(\eta) = \sum_{i=1}^{N_c} \left(\sum_{j =0}^k  \phi_{i,j}(\eta) \mh_{1,j}^i + \sum_{j =0}^{k-1} \psi_{i,j}(\eta) (\mh_{2,j}^{i})^{\perp}\right),
\end{equation}
where $\mg_{2,j}^i=n(\mg_{1,j+1}^i-\mg_{1,j}^i)$, and
\begin{equation}\label{equ:basis_inte}
\begin{aligned}
   & \phi_{i,j}(\eta)=\int\limits_0^1\nabla_1 G_1(\mc_i(t),\eta)\cdot \mc_i'(t)^{\perp}B^n_j(t)\,dt,\\
   &\psi_{i,j}(\eta)=\int\limits_0^1 G_1(\mc_i(t),\eta) B^n_j(t)\,dt,\\
   &\tilde{\phi}_{i,j}(\eta)=\int\limits_0^1\nabla_1 G_2(\mc_i(t),\eta)\cdot \mc_i'(t)^{\perp}B^n_j(t)\,dt,\\
   &\tilde{\psi}_{i,j}(\eta)=\int\limits_0^1 G_2(\mc_i(t),\eta) \|\mc_i'(t)\|^2 B^n_j(t)\,dt,
 \end{aligned}
\end{equation}
with $\nabla_1 G_1(\xi, \eta) = \frac{\xi - \eta}{2\pi \|\xi - \eta\|^2}$ and $\nabla_1 G_2(\xi, \eta) = \frac{(\xi - \eta)(2 \log(\|\xi-\eta\|)-1)}{8\pi} $.
\paragraph{Analytical computation of the integrals and their derivatives}
The terms related to \(\phi_{i,j}\) and \(\psi_{i,j}\) correspond precisely to the polynomial Green coordinates, which have already been computed by~\cite{MichelThiery2023} for polygonal cages and by~\cite{liu2024polygc} for high-order cages. Here, we first present the computation methods for \(\tilde{\phi}_{i,j}\) and \(\tilde{\psi}_{i,j}\), followed by the derivation of their derivatives.

For \(\tilde{\phi}_{i,j}\), note that \(\nabla_1 G_2(\mc_i(t),\eta) \cdot \mc_i'(t)^{\perp} = (-\eta \cdot \mc_i'(t)^{\perp})(2 \log(\| c(t) - \eta\|) - 1)\). The integral term in \(\tilde{\phi}_{i,j}\) is expressed as the product of a polynomial function and a logarithmic term. For simplicity, we can focus on calculating the following terms:
\begin{equation}\label{equ:cal_termsingle}
\begin{aligned}
    &\int_0^1 t^j \log(\|\mc_i(t) - \eta \|)\, dt \\
    &= \frac{\log(\|\mc^i_{m} - \eta\|)}{j+1} + \frac{1}{j+1}\int_0^1 \frac{t^{j+1}(c(t)-\eta)\cdot c'(t)}{\|c(t)-\eta\|^2}\, dt
\end{aligned}
\end{equation}
where \(j = 0, \dots, n\). The problem is transformed into computing $F_{h,2}^{c,\eta}=\int_0^1\frac{t^h}{\|c(t)-\eta \|^2}\,dt, h=j+1,\dots,j+2m$. 
Similarly, the integral term in \(\tilde{\psi}_{i,j}\) can also be expressed as a product of a polynomial and a logarithmic term. It can be calculated in the same manner as \(\tilde{\phi}_{i,j}\), but with a higher-degree polynomial, where \(j = 0, \dots, 4m + n - 2\).

The derivative follows a similar form to the integral. For \(\tilde{\phi}_{i,j}'\), the non-trivial terms only involve computing the derivative of \eqref{equ:cal_termsingle}. In fact, for the derivatives of the four integrals in \eqref{equ:basis_inte}, the non-trivial terms only include:  
$
F_{h,4}^{c,\eta} := \int_0^1 \frac{t^h}{\|c(t) - \eta \|^4} \, dt.
$ 

Thus, it suffices to calculate \( F_{h,2}^{c,\eta} \) and \( F_{h,4}^{c,\eta} \). \cite{liu2024polygc} provide the following lemma:  
\begin{lemma}  
For \( p(t) = \prod_{i=1}^{n} (t - \omega_i)^{n_i} \), the following holds:  
\[
\int_0^1 \frac{t^h}{p(t)} dt = \sum_{i=1}^{n} \text{Res}\left(\frac{w^m \left( \log\left(1 - \frac{1}{w}\right) + \sum_{k=1}^h \frac{1}{k w^k} \right)}{p(w)}, \omega_i\right).
\]  
\end{lemma}  
We apply this lemma to compute \( F_{h,2}^{c,\eta} \) and \( F_{h,4}^{c,\eta} \) by setting $p(t)=\|c(t) - \eta \|^2$ and $p(t)=\|c(t) - \eta \|^4$. This involves solving the equation \(\|c(t) - \eta \|^2 = 0\), which has an analytical solution when the degree of \(c(t)\) is less than 5.

\subsubsection{Solving boundary integral equations}
\paragraph{Singular integrals}
To obtain the system of algebraic equations, we sample \(\eta^{\text{b}}\) on the boundary. 
At this stage, evaluating the integrals above involves the computation of singular integrals, which are extensively addressed in the context of boundary element methods~\cite{Fischer1999ApplicationOH}. We follow the suggestion in~\cite{Weber2012} that the limits of these integrals exist as \(\eta\) approaches the boundary. In practice, we slightly displace $\eta^{\text{b}}$ inward towards the cage and compute the integrals directly. This approach has yielded accurate results in our experiments.

\paragraph{Matrix constraints}
Substituting \(\eta^{\text{b}}\) on the boundary into \eqref{equ:dis_Int1} and \eqref{equ:dis_Int2} yields the following matrix constraints:

\begin{equation}\label{equ:matrix1}
    M_nG_1=\Phi_nG_1+\Psi_n G_2+\bar{\Phi}_k H_1+\bar{\Psi}_k H_2,
\end{equation}
\begin{equation}\label{equ:matrix2}
    M_k H_1=\Phi_k H_1+\Psi_k H_2.
\end{equation}
Here, \( G_1 \) (or \( G_2 \)) is an \( N_c \cdot n \)-dimensional column vector composed of \( \mg_{1,j}^i \) (or \( s_i (\mg_{2,j}^{i})^{\perp} \)), and \( H_1 \) (or $H_2$) is an \( N_c\cdot k \)-dimensional column vector composed of \( \mh_{1,j}^i \) (or \( (\mh_{2,j}^{i})^{\perp} \)). Note that, due to continuity, adjacent edges share a common \( \mg_{1,0}^i \) and \( \mh_{1,0}^i \) (or \( \mg_{1,n}^i \) and \( \mh_{1,k}^i \)). The \(l\)-th row of the matrices in \eqref{equ:matrix1} and \eqref{equ:matrix2} corresponds to the calculations of \eqref{equ:dis_Int1} and \eqref{equ:dis_Int2} at the \(l\)-th sampling point. \(M_n\) and \(M_k\) collect the coefficients of the \(n\)-th and \(k\)-th degree Bernstein basis functions at $\eta^{\text{b}}$ located on the corresponding boundary element. $\Phi_n$ collects $\phi_{i,j}(\eta^{\text{b}})$ in \eqref{equ:dis_Int1} in the order of $G_1$. $\Psi_n,\bar{\Phi}_k,\bar{\Psi}_k, \Phi_k$ and $\Psi_k$ are defined in the same way.

\begin{figure}[t]
  \centering
  \begin{overpic}[width=1.0\linewidth]{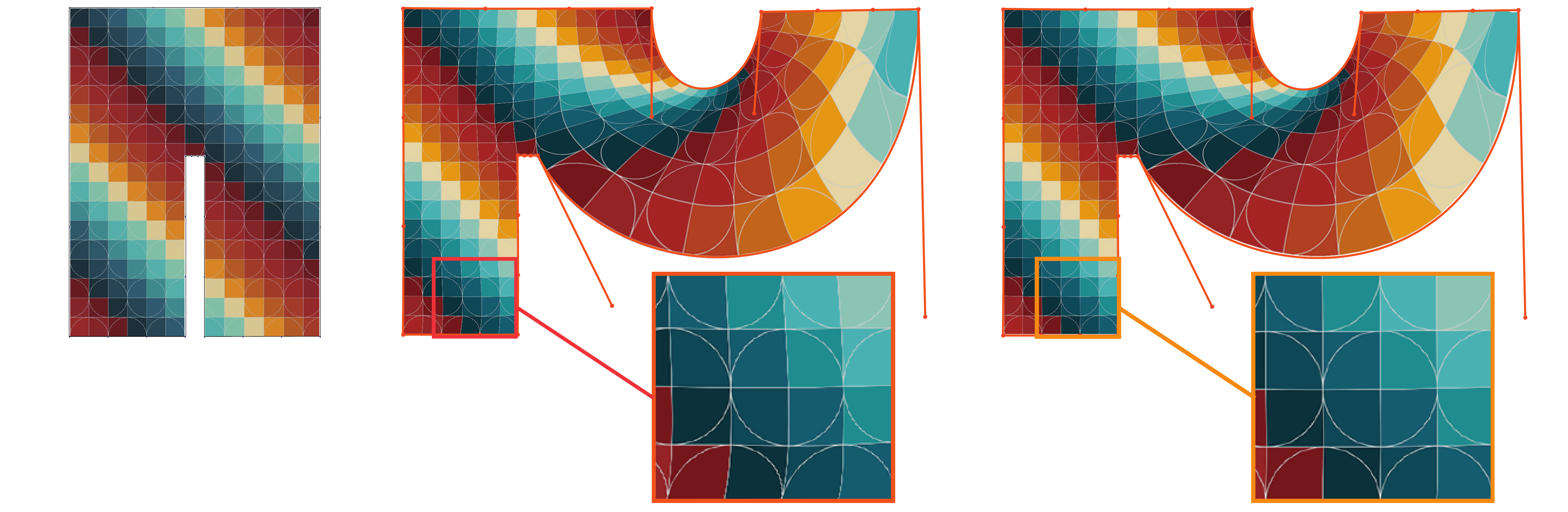}
    {
    \put(7,-1.5){\small Rest pose}
    \put(40,-1.5){\small $\text{BiHC}_{1}$}
 \put(78,-1.5){\small $\text{BiHC}_{1,2}$}
    }
  \end{overpic}
  \caption{
  Comparison of the two methods for solving the algebraic equations. When only the right side of the pants is deformed, the result of $\text{BiHC}_{1}$ introduces perturbations in the left leg area, whereas $\text{BiHC}_{1,2}$ maintains a disturbance-free deformation.
  }
  \label{fig:linear-solving}
\end{figure}

\paragraph{Approximating algebraic equations}
Two methods are used to solve the system of algebraic equations: one strictly enforces the Laplacian constraints, while the other applies them more leniently. For simplicity, we only consider the case \(k = n\), as the case \(k > n\) is analogous. At this point, the eight matrices in the system of equations are simplified to four. They lead to the following expressions for our regularized biharmonic coordinates:
\begin{equation}\label{equ:final_BHC}
    \begin{aligned}
f(\eta) &= \big[ \phi(\eta) + (\bar{\phi}(\eta) C_L + \bar{\psi}(\eta) C_D)(M_n - \Phi_n) \big] G_1 \\
&\quad + \big[ \psi(\eta) - (\bar{\phi}(\eta)) C_L + \bar{\psi}(\eta) C_D)\Psi_n \big] G_2 . \\
&:=\alpha(\eta)G_1+\beta(\eta)G_2
\end{aligned}
\end{equation}
Here, \(\phi(\eta)\) represents a row vector composed of the evaluations of \(\phi_{i,j}\) at \(\eta\). \(\bar{\phi}(\eta)\), \(\bar{\psi}(\eta)\), and \(\psi(\eta)\) are defined in the same way. \(\alpha(\eta)\) and \(\beta(\eta)\) together form our biharmonic coordinates.
In the first method proposed by~\cite{Weber2012}, denoted as $\text{BiHC}_1$, they optimize only \eqref{equ:dis_Int1} while strictly adhering to \eqref{equ:dis_Int2}. The matrices are given as \(C_L = (\bar{\Phi}_n + \bar{\Psi}_n A)^{-1}\) and \(C_D = A (\bar{\Phi}_n + \bar{\Psi}_n A)^{-1}\), where \(A = \Psi_n^{-1} (M_n - \Phi_n)\). 
In the second method by~\cite{thiery2024biharmonic}, denoted as $\text{BiHC}_{1,2}$, they simultaneously optimize \eqref{equ:dis_Int1} and \eqref{equ:dis_Int2}. The matrices are determined using 
\[
\begin{pmatrix} 
C_L \\ 
C_D 
\end{pmatrix} = \left( E^T E + F^T F \right)^{-1} E^T,
\]
where \(E = (\bar{\Phi}_n; \bar{\Psi}_n)\) and \(F = (\Phi_n - M_n; \Psi_n)\).
Our experiments find that $\text{BiHC}_{1}$ and $\text{BiHC}_{1,2}$ yield comparable results when the magnitude of deformation is minimal. However, when confronted with more challenging deformations, $\text{BiHC}_{1,2}$ demonstrates superior deformation results compared to $\text{BiHC}_{1}$ (see Fig.~\ref{fig:linear-solving}). This observation aligns with the insights gained from \cite{thiery2024biharmonic}: the joint optimization of Laplacian and Dirichlet constraints allows for increased degrees of freedom.


\subsection{Deformation controls}\label{sec:deformation}

\paragraph{Improving accuracy}
To improve the accuracy of numerical solutions, three methods can be used:
\begin{enumerate}
    \item Increase the number of boundary elements to expand the discretized approximation space. \item Increasing the degree of \(k\), similar to Method (1), results in a larger approximation space with greater degrees of freedom. 
    \item Increase the number of sampling points \( \eta^\text{b} \) to form an over-constrained system of equations, which enforces the integral equations to hold at more points in the least-squares sense, thereby making the deformation results appear smoother. 
\end{enumerate}
Our experiments find that Methods (1) and (2) yield comparable results.
In practice, each edge of the cage is subdivided into four elements, and we uniformly sample \(2 n\) points on the domain of each element to construct the equation system.
%

\paragraph{Deformation energies}
In~\eqref{equ:g}, different choices of \( s_j \) correspond to different Neumann boundary conditions. 
Setting $s_j$ simply to 1 has yielded satisfactory results in many examples, as they exhibit conformal behavior near the boundary, similar to Green coordinates.
However, for certain cage deformations that deviate significantly from conformality, this local conformality near the boundary can result in larger shear within the interior of the cage.
In such cases, it becomes necessary to consider the global deformation of the cage.
Following the approach outlined in~\cite{Weber2012} and \cite{thiery2024biharmonic}, we present two methods for automatically deducing boundary derivatives from the given boundary positions. 
Since \(s_i < 0\) can lead to degeneracy and folding near the boundaries during deformation, we enforce \(s_i > 0\).
We sample point constraints $C = \{q_k\}$ on $\partial \Omega$ to discretize our two different deformation energies:
\begin{enumerate}
\item As-Harmonic-As-Possible (AHAP) energy: $\sum_k \|\Delta f(q_k)\|^2$. We compute this energy using~\eqref{equ:dis_Int2}. 
\item As-Affine-As-Possible (AAAP) energy: $\sum_k \|f'(q_k)\|^2$. We compute such energy via~\eqref{equ:final_BHC}. Its evaluation requires derivatives of the integrals $\phi(\eta),\psi(\eta),\bar{\phi}(\eta),\bar{\phi}(\eta)$, which we provided the computation in Section \ref{sec:Reformulating boundary integral equations}. 
\end{enumerate}
We find that the unit normal derivative, along with the normal derivatives of the AHAP and AAAP energies, progressively enhances the internal conformality. All three methods are capable of producing viable deformations, depending on the application scenario.

\begin{figure}[t]
  \centering
  \begin{overpic}[width=1.0\linewidth]{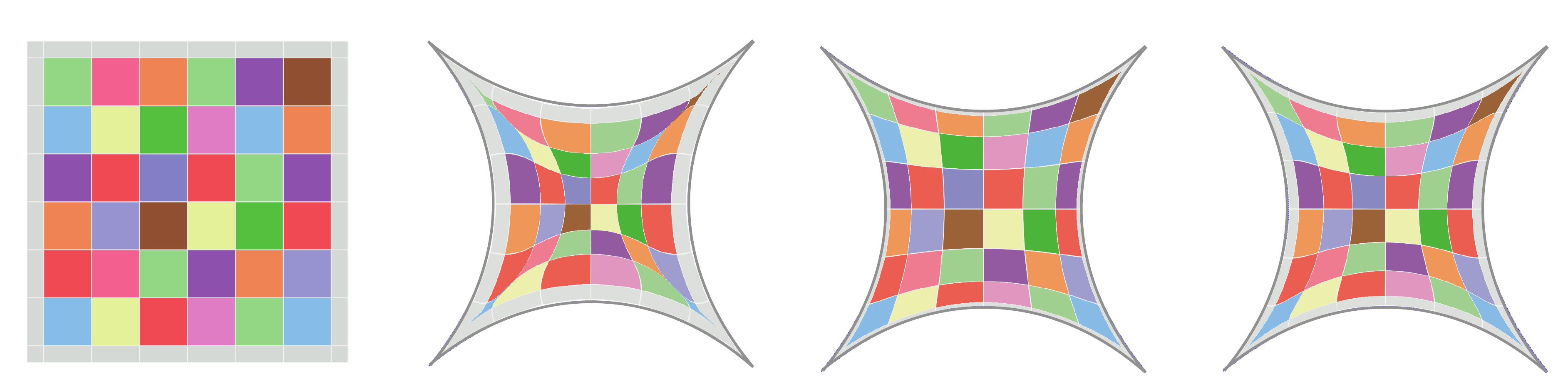}
    {
    \put(6,-1.5){\small Rest pose}
    \put(35,-1.5){\small Unit}
    \put(58,-1.5){\small AHAP}
 \put(85,-1.5){\small AAAP}
    }
  \end{overpic}
  \caption{
  Comparison of the deformations using different scaling factors, including unit and optimized ones through minimizing deformation energies.
  }
  \label{fig:scaling}
\end{figure}

\begin{figure}[t]
  \centering
  \begin{overpic}[width=0.99\linewidth]{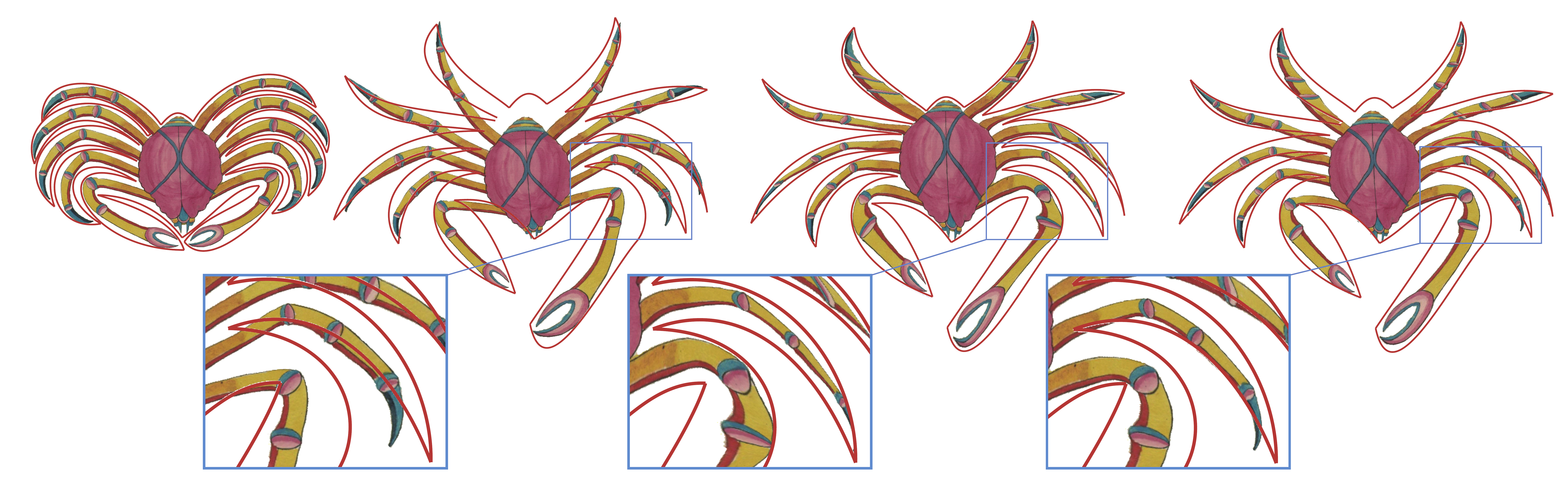}
    {
    \put(2,-1.5){\small \textbf{(a)} Rest pose}
     \put(26,-1.5){\small \textbf{(b)} $\omega=0$}
      \put(55,-1.5){\small \textbf{(c)} $\omega=1$}
 \put(80,-1.5){\small \textbf{(d)} $\omega=0.5$}
    }
  \end{overpic}
  \caption{
  The result of the weighted coordinates (d), which are expressed as a linear combination of Green coordinates (b) and biharmonic coordinates (c).
  }
  \label{fig:weights}
\end{figure}
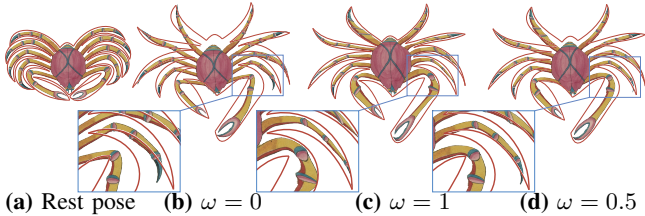

\paragraph{Different weights}
Conformal deformation cannot be achieved using Biharmonic coordinates unless the cage deformations are already conformal, as conformality and interpolation are inherently conflicting properties. However, we can use Biharmonic coordinates to allow users to balance and prioritize between the two aspects.
In \eqref{equ:final_BHC}, one can divide the Biharmonic coordinates into two components: \(\{\alpha_c,\beta_c\}\) and \(\{\alpha_{\text{i}},\beta_{\text{i}}\}\), which are defined as follows. 
\begin{equation*}
    \begin{split}
    &\alpha_c(\eta)=\phi(\eta),\\
    &\beta_c(\eta)=\psi(\eta),\\
     &\alpha_{\text{i}}(\eta)=(\bar{\phi}(\eta) C_L + \bar{\psi}(\eta) C_D)(M_n - \Phi_n),\\
     &\beta_{\text{i}}(\eta)=- (\bar{\phi}(\eta)) C_L + \bar{\psi}(\eta) C_D)\Psi_n.
    \end{split}
\end{equation*}
where \(\{\alpha_c,\beta_c\}\) represents Green coordinates that provide the conformal deformation, and \(\{\alpha_{\text{i}},\beta_{\text{i}}\}\) is a biharmonic function that pulls the boundaries of the conformal deformation back to the cage boundaries. 
We define a new coordinate as \(\{\alpha_c + w \alpha_{\text{i}},\beta_c+w\beta_{\text{i}}\}\), where users can adjust the value of \(w\) to control the deformations. 
When \(w = 0\), the deformation is conformal; when \(w = 1\), the result fully interpolates the boundaries. 
Typically, choosing an intermediate value for \(w\) yields a more intuitive and satisfactory deformation result (see Fig.~\ref{fig:weights}).

%% file: src/results.tex
\section{Experiments and evaluations}\label{sec:results}


\begin{figure}[t]
  \centering
  \begin{overpic}[width=0.99\linewidth]{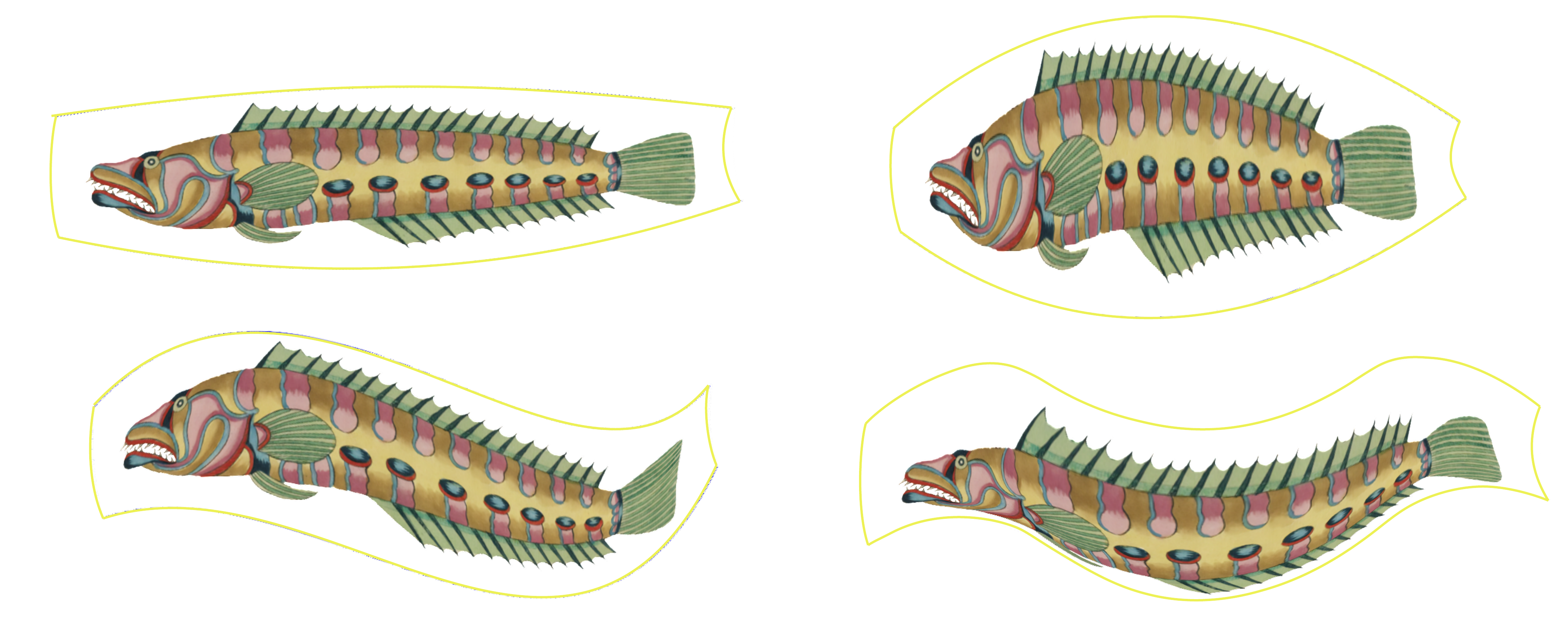}
    {
    }
  \end{overpic}
  \vspace{-2mm}
  \caption{
The input quadratic cage (top left) is deformed into cages of different orders: quadratic (top right), cubic (bottom left), and quartic (bottom right).
  }
  \label{fig:order}
\end{figure}

\begin{figure*}[t]
  \centering
  \begin{overpic}[width=0.99\linewidth]{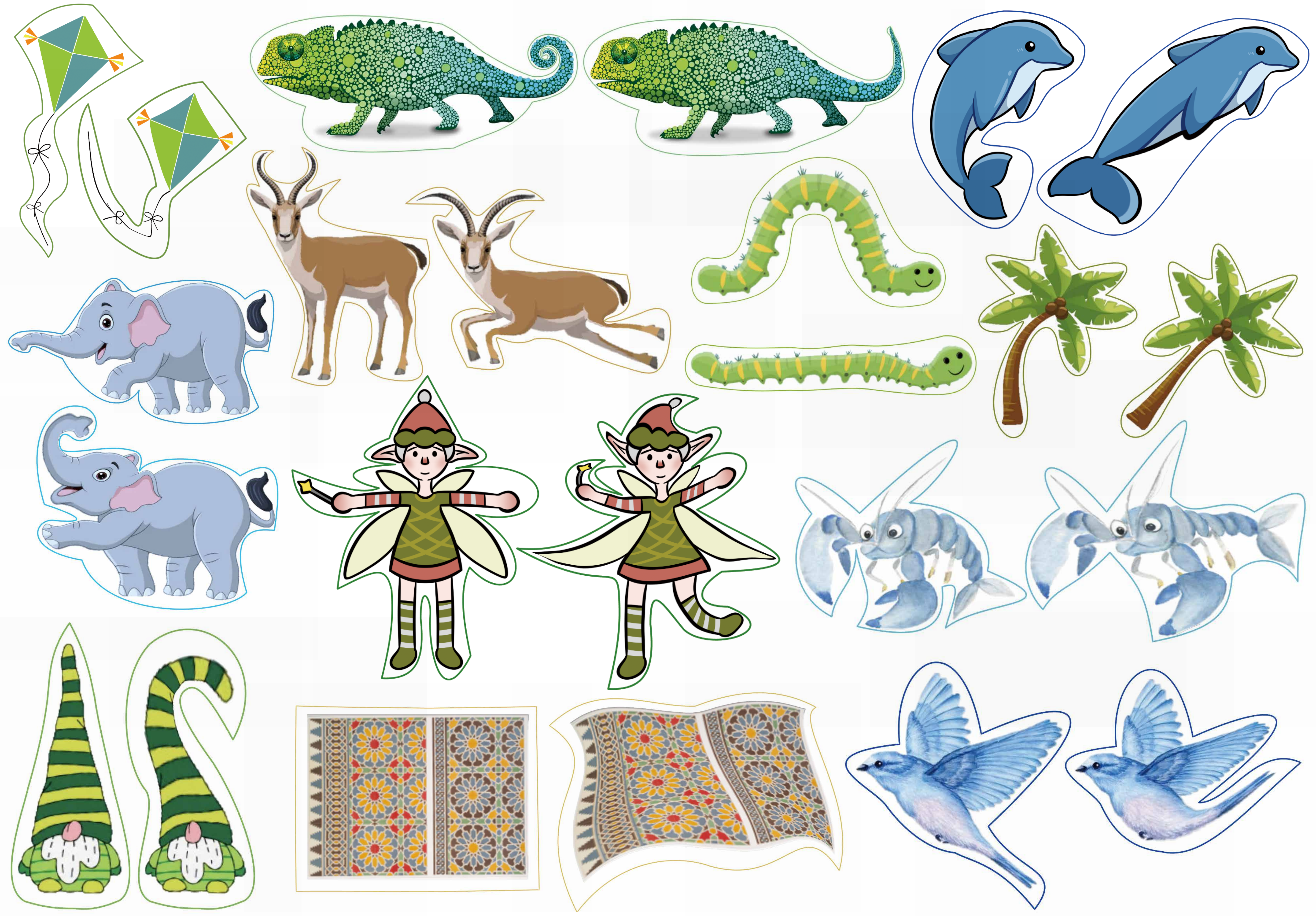}
    {
    }
  \end{overpic}
  \vspace{-3mm}
  \caption{
  Applying our method to more cages with polynomial curves of different degrees.
  }
  \label{fig:more-examples}
\end{figure*}

\paragraph{Different target orders}
When setting the boundary conditions \eqref{equ:g}, selecting different values of \(n\) corresponds to deforming the cage to different orders, as shown in Fig. \ref{fig:order} and Fig. \ref{fig:more-examples}, regardless of whether the initial cage is linear or of a higher order. Although we allow the deformed cage to have an arbitrarily high order, it is important to note that higher-order cages result in more complex integral calculations \eqref{equ:basis_inte} and larger matrix constraints for solving the coordinates \eqref{equ:final_BHC}, leading to increased precomputation time.


\paragraph{Comparisons}
We select the Cubic MVC \cite{Li2013} and the Polynomial Green Coordinate (PolyGC) \cite{MichelThiery2023} as the competitors when the rest cage is linear, and choose \cite{liu2024polygc} and \cite{Lin2024PolynomialCC} when the rest cage is of higher order.

\begin{figure*}[t]
  \centering
  \begin{overpic}[width=0.99\linewidth]{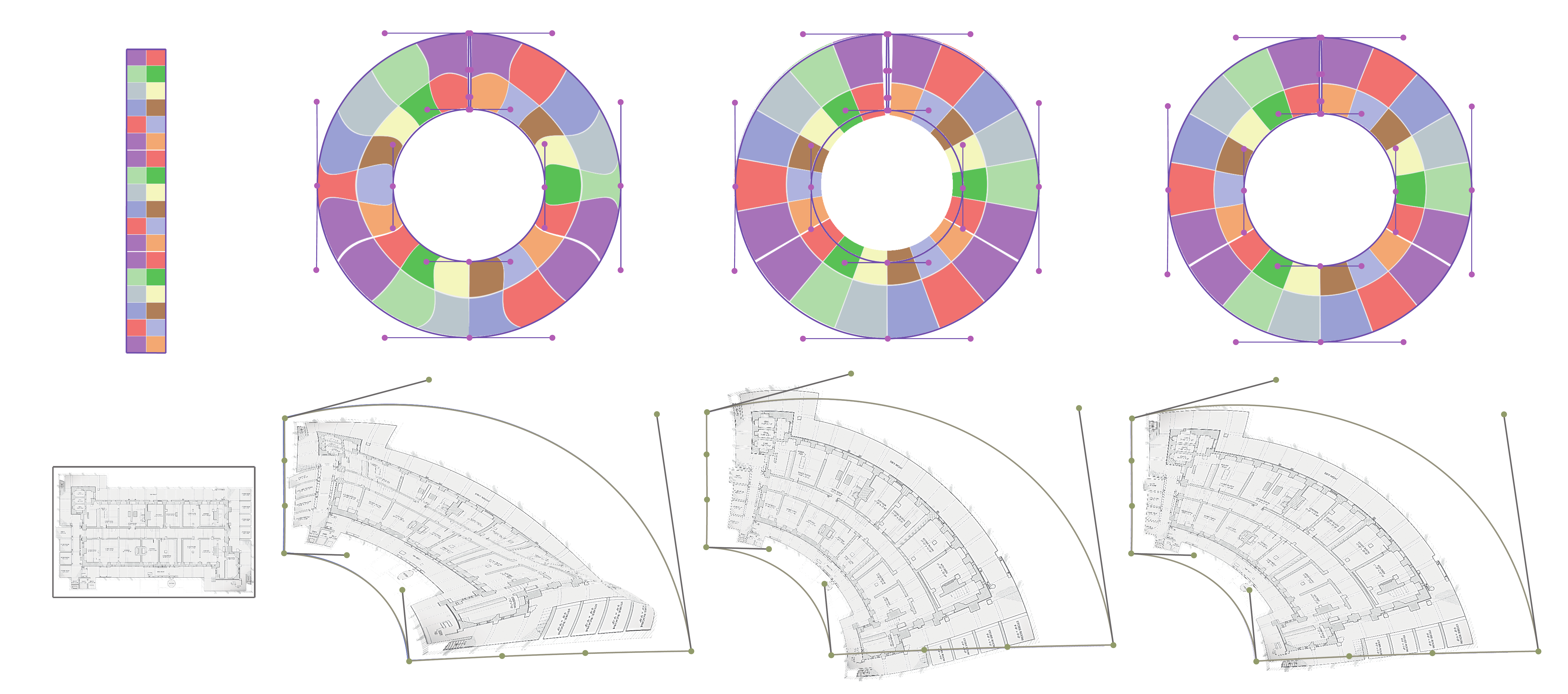}
    {
\put(7,-0.5){\small Rest pose}
 \put(28,-0.5){\small Cubic MVC}
 \put(55,-0.5){\small Poly GC}
 \put(84,-0.5){\small Ours}
    }
  \end{overpic}
  \vspace{-0mm}
  \caption{
  Comparisons with Cubic MVC and PolyGC on two examples using the same polygonal cages. At the bottom, we use weighted coordinates with $w=0.5$.
  }
  \label{fig:more-cmp}
\end{figure*}

\begin{figure*}[t]
  \centering
  \begin{overpic}[width=1.0\linewidth]{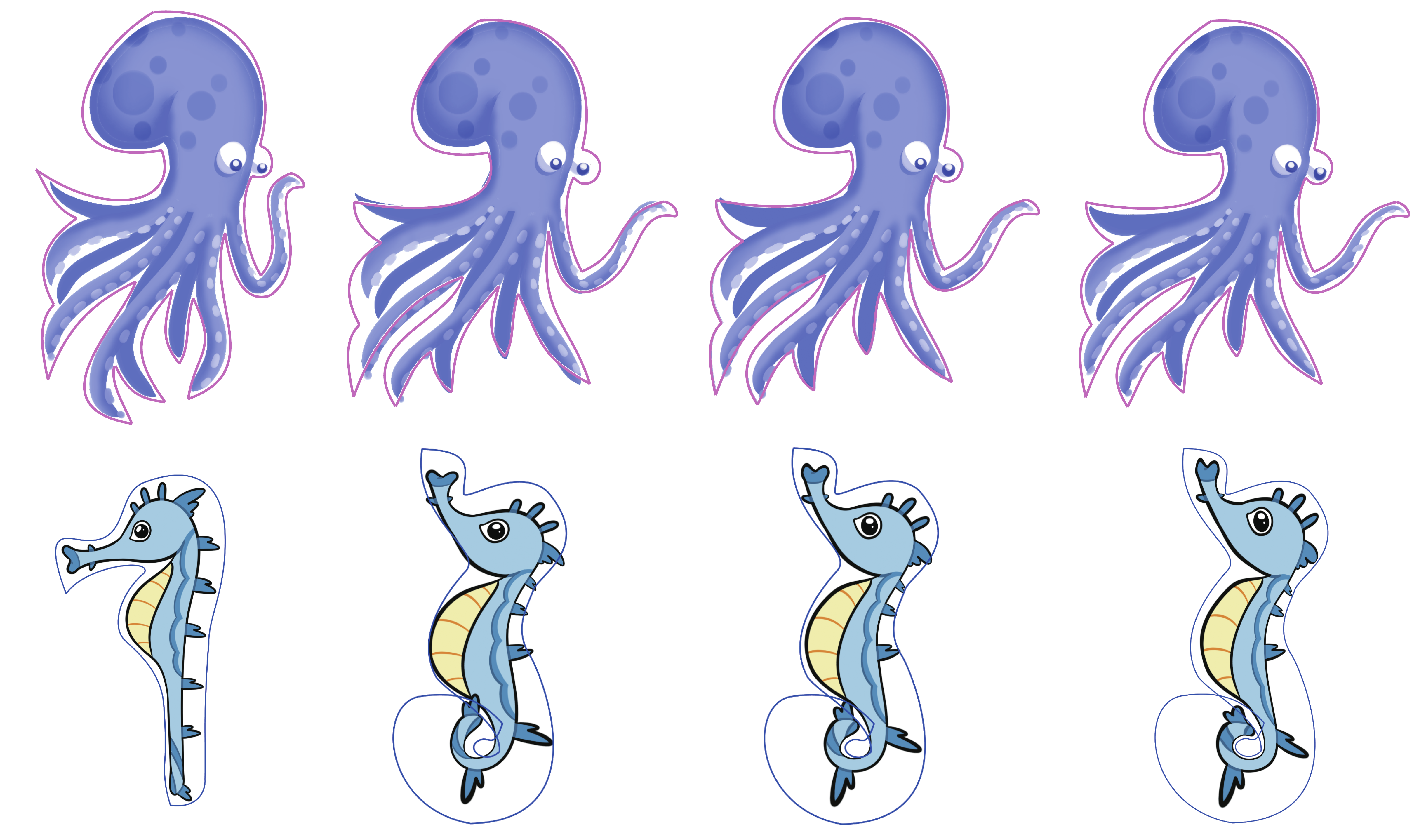}
    {
 \put(10,-0.5){\small Rest pose}
 \put(28,-0.5){\small $w=0$ (CurvedGC)}
 \put(57,-0.5){\small $w=0.5$}
 \put(84,-0.5){\small $w=1$}
    }
  \end{overpic}
  \caption{
   Comparisons with CurvedGC on two examples using the same cubic cages.
  }
  \label{fig:more-curvedcmp}
\end{figure*}

Like our coordinates, the Cubic MVC interpolates both the functions and their derivatives on the boundary. However, these coordinates suffer from the same artifacts that regular mean-value coordinates do. Namely, they perform poorly on non-convex shapes (see Fig.~\ref{fig:cmp-cmvc}). Furthermore, our coordinates inherit the advantages of harmonic coordinates, minimizing the so-called Hessian energy and possessing a harmonic Laplacian, which results in deformations with less distortion (see Fig.~\ref{fig:more-cmp}). Additionally, the Cubic MVC lacks support for higher-order cages.

Combining the polynomial Cauchy coordinates with their inverse mapping, \cite{Lin2024PolynomialCC} enables deformation between curved cages. However, their method requires the user to specify an additional intermediate linear cage, and the inverse mapping computation relies on numerical integration, lacking an analytical expression.

Our coordinates are more closely related to PolyGC and CurvedGC, both in terms of the coordinate expressions and the deformation results. 
PolyGC and CurvedGC trade the interpolation property for the conformal one. Our coordinates provide a flexible approach for a trade-off between conformality and interpolation. As shown in Figs.~\ref{fig:weights} and \ref{fig:more-curvedcmp}, we apply a simple linear weighting to the coordinates of all points. More complex weighting schemes can also be used, such as using a distance function as a weight, allowing the user to specify regions within the cage that prioritize either conformality or interpolation.

%% file: src/conclusion.tex
\section{Conclusion and Discussion}\label{sec:conclusion}
We propose polynomial 2D biharmonic coordinates for closed high-order cages containing polynomial curves of any order. Our coordinates are obtained by extending the classical
2D biharmonic coordinates using high-order BEM, enabling the transformation between
polynomial curves of any order. When applying our coordinate
to 2D cage-based deformation, users manipulate the \Bezier
control points to quickly generate the desired conformal deformation. We demonstrate the effectiveness and practicability
of our coordinates by extensively testing them on various 2D deformations.

\paragraph{limitation and feature work}
Our coordinates are defined in 2D. Nevertheless, they can be derived in 3D (although closed-form expressions might be more difficult to obtain). In computer graphics, 3D models are more widely used, and it is exciting to generalize our coordinates to 3D.

Our approach requires setting all deformation curves by hand. Designing variational frameworks such as \cite{Weber2012,Lin2024PolynomialCC} to allow artists to deform shapes with very few control points is an interesting avenue for feature work.


%% file: src/acknowledgments.tex
\section*{Acknowledgments}\label{sec:acknowledgments}
We would like to thank the anonymous reviewers for their constructive suggestions and comments.
This work is partially supported by the National Natural Science Foundation of China (62272429, 62102355), the Major Project of Science and Technology of Anhui Province (202203a05020050), and the National Key R\&D Program of China (2022YFB3303400). 